\documentclass{PoS}
\usepackage{amssymb,fontenc,times,mathptmx,graphicx}
\usepackage{amsmath}
\usepackage[table]{xcolor}
\usepackage{cellspace} 
\title{Cosmological scalar fields and Big-Bang nucleosynthesis}
\ShortTitle{Cosmological scalar fields and Big-Bang nucleosynthesis}
\author{\speaker{Jean-Fran\c{c}ois Coupechoux}\\
        Univ Lyon, Univ Claude Bernard Lyon 1, CNRS/IN2P3, IP2I Lyon, UMR 5822, F-69622, Villeurbanne, France\\
        E-mail: \email{j-f.coupechoux@ipnl.in2p3.fr} }
\author{Alexandre Arbey \thanks{Also Institut Universitaire de France, 103 boulevard Saint-Michel, 75005 Paris, France}\\
        Univ Lyon, Univ Claude Bernard Lyon 1, CNRS/IN2P3, IP2I Lyon, UMR 5822, F-69622, Villeurbanne, France\\
        E-mail: \email{alexandre.arbey@ens-lyon.fr} }

\abstract{The nature of dark matter and of dark energy which constitute more than $95\%$ of the energy in the Universe remains a great and unresolved question in cosmology. Cold dark matter can be made of an ultralight scalar field dominated by its mass term which interacts only gravitationally. The cosmological constant introduced to explain the recent acceleration of the Universe expansion can be easily replaced by a scalar field dominated by its potential. More generally, scalar fields are ubiquitous in cosmology: inflaton, dilatons, moduli, quintessence, fuzzy dark matter, dark fluid, etc. are some examples. One can wonder whether all these scalar fields are independent. The dark fluid model aims at unifying quintessence and fuzzy dark matter models with a unique scalar field. One step futher is to unify the dark fluid model with inflation. In the very early Universe such scalar fields are not strongly constrained by direct observations, but Big-Bang nucleosynthesis set constraints on scalar field models which lead to a modification on the abundance of the elements. In this talk we will present a scalar field model unifying dark matter, dark energy and inflation, and study constraints from Big-Bang nucleosynthesis on primordial scalar fields.}

\FullConference{40th International Conference on High Energy Physics\\Prague, Czech Republic\\28 July - 6 August 2020}

\begin{document}


\section{Introduction}
Scalar fields are ubiquitous in cosmology. Fuzzy dark matter model \cite{Hu:2000ke} has for example been introduced to replace cold dark matter with a scalar field dominated by its mass term, and such a scalar field behaves like collisionless matter. Quintessence models \cite{Zlatev:1998tr} on the other hand replace the cosmological constant with a scalar field. The energy density of these models evolves with time and may have played a role at earlier stages of the Universe. Inflation can also be described with a scalar field. 

One possibility to reduce the number of different scalar fields involved in cosmology is to unify them. In the first section, we will present the dark fluid model \cite{Peebles:2000yy,Arbey:2006it} which describes both dark energy and dark matter with a single scalar field. In the second section, we will go further by introducing a scalar field to rule them all, i.e. unifying inflation and dark fluid. In the final section, we will derive constraints on scalar field scenarios from Big-Bang nucleosynthesis, since the presence of scalar fields can affect the observed abundance of the elements. 

\section{Dark Fluid model}

\begin{figure}[h!]\centering
\includegraphics[width=11cm]{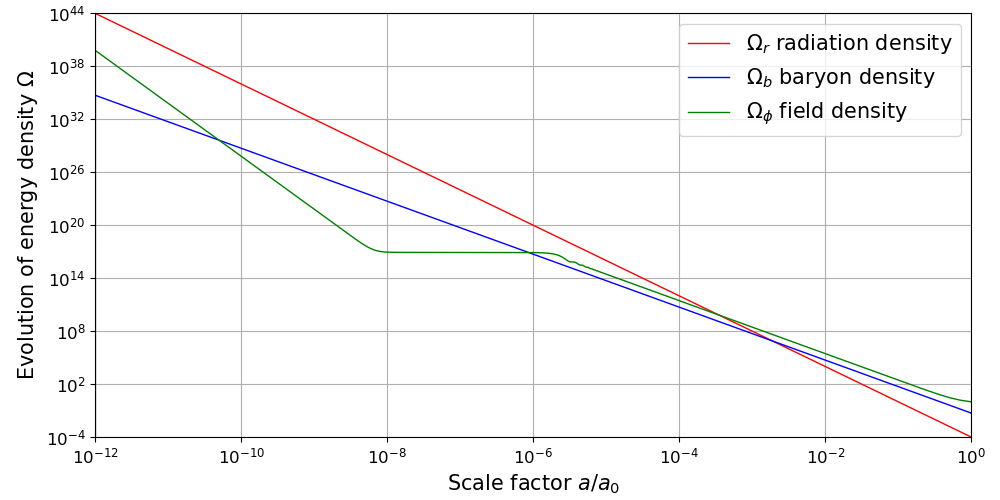}
\caption{Evolution of the dark fluid scalar field density (green), radiation density (red) and baryon density (blue) fractions as functions of the scale factor.\label{fig::cosmo_evo}}
\end{figure}

The dark fluid model aims at unifying dark energy and dark matter with a single scalar field. To reproduce a cold dark matter behaviour, the scalar field has to oscillate quickly around the minimum of its potential. Its value at the minimum needs to be nonzero in order to create an acceleration of the expansion, as explained by the cosmological constant in the $\Lambda$CDM model. The following system of equations gives the cosmological evolution for an isotropic and homogeneous Universe described by the Robertson and Walker metric and the Klein-Gordon equation which governs the scalar field evolution:  
\begin{equation}
\begin{aligned}
	& H^2 = \frac{8\pi G}{3} \left( \rho_{\phi} + \rho_r + \rho_m \right)\,, \\
	& 2\dot{H} +3H^2 = -8\pi G \left( P_{\phi} + P_r + P_m \right)\,, \\
	& \ddot{\phi} + 3H\dot{\phi} + \frac{dU}{d\phi} = 0\,.
\end{aligned}
\end{equation}
The radiation energy density $\rho_r$ evolves according to $a^{-4}$ and the baryonnic matter energy density $\rho_m$ evolves according to $a^{-3}$ where $a$ is the scale factor. Both energy densities are drawn in Figure~\ref{fig::cosmo_evo} and are the same as in the $\Lambda$CDM model. To determine the density of the scalar field $\rho_{\phi}$, one needs to define the potential $U$, but its shape is still an open question. The simplest potential can be defined with the two parameters $V_0$ and $m$:
\begin{equation}
U(\phi) = V_0 + \frac{1}{2}m^2 \phi^2\,.
\end{equation}
The constant $V_0=\Lambda c^4 / 8\pi G$ leads to a dark energy behaviour with $\Lambda$ the cosmological constant, and the mass term $m$ leads to a dark matter behaviour. The value of $m$ is approximately equal to $10^{-22}$~eV, which corresponds to the mass of the fuzzy dark matter model. At galactic scale, the scalar field forms Bose-Einstein condensates, which may constitute galaxy-sized dark matter halos. For a typical halo of $10$~kpc, the Compton wavelength $l=h/mc$ requires such a tiny mass $m$. With this value, one can reproduce the flat spiral galaxy rotation curves \cite{Arbey:2003sj} and avoid the cuspy halo and missing satellite problems~\cite{Hui:2016ltb}. 

The evolution of the energy density $\rho_{\phi}$ is also drawn in Figure \ref{fig::cosmo_evo}: when the scalar field evolution is dominated by its kinetic energy, the density decreases as $a^{-6}$. Later the potential is no longer negligible and one can observe a plateau. When the scalar field oscillates quickly around its minimum, the energy density decays as $a^{-3}$, and more recently the constant $V_0$ leads to an accelerated expansion of the Universe. There in the dark fluid model, a single scalar field can replace both dark matter and dark energy simultaneously. 

\section{Triple Unification}
Several models of triple unification have already been studied in the literature, but these models do not explain dark matter as in the dark fluid model, i.e. with a scalar field with an ultralight mass term $m\sim10^{-22}$ eV. In this section we present a more natural triple unification scenario by assuming a non-minimal coupling to the gravity $\phi^2R^2$ and a symmetry breaking before inflation (see Ref. \cite{Arbey:2020ldf}). Considering a $Z_2$ symmetry for the scalar field and a quadratic coupling in $R$, the model is defined by the action:
\begin{equation}
\mathcal{S}=\int d^4x \sqrt{-g}\left[\frac{1}{2\tilde{\kappa}^2}\left(R+\alpha \phi^2 R + \beta \phi^2 R^2 \right)-\frac{1}{2}g^{\mu \nu}\partial_{\mu}\phi\partial_{\nu}\phi-V(\phi) \right]\,,
\label{action_sym}
\end{equation}
where $\tilde{\kappa}$ is a modified Einstein's constant and $V$ the potential of the scalar field defined by:
\begin{equation}
V(\phi) = V_0 + \frac{m^2}{8v^2}\left(\phi^2-v^2\right)^2\,.
\end{equation}

For $\phi$ equal to zero, the potential has a local maximum, around which the theory is unstable. The two minima correspond to $\phi = \pm v$. When the scalar field goes to one of these minima the $Z_2$ symmetry is spontaneously broken. By replacing $\phi$ by $\xi \pm v$ where $\xi$ characterizes the variation of the scalar field around the minimum, the action \eqref{action_sym} becomes:
\begin{equation}
\mathcal{S}=\int d^4x \sqrt{-g}\left[\frac{1}{2\kappa^2}\left(R+\frac{\beta v^2}{\alpha v^2 + 1/(1 \pm 2\xi /v + \xi^2/v^2)} R^2 \right)-\frac{1}{2}g^{\mu \nu}\partial_{\mu}\xi\partial_{\nu}\xi- V(\xi) \right]\,,
\label{action_break}
\end{equation}
with:
\begin{equation}
\begin{aligned}
&\kappa = \frac{\tilde{\kappa}}{\sqrt{1 + \alpha v^2 (1 \pm 2\xi /v + \xi^2 /v^2)}}\,,\\
&V(\xi) = V_0 + \frac{m^2}{2}\xi^2 \pm \frac{m^2}{2v}\xi^3 + \frac{m^2}{8v^2}\xi^4\,.
\end{aligned}
\label{pot}
\end{equation}
In the $|\xi| \ll v$ limit, $\kappa$ is constant and equal to the Einstein's constant. If one neglects the scalar field variation $\xi$, only the $R$ and $R^2$ terms have an impact on the Universe evolution after the symmetry breaking. The $R^2$ term produces an inflationary period which will be similar to the one of Starobinsky inflation \cite{Starobinsky:1980te}:
\begin{equation}
\mathcal{S}=\int d^4x \sqrt{-g}\Big[\frac{1}{2\kappa^2} \left(R+\frac{\beta v^2}{(1+\alpha v^2)M_P^2} R^2 \right) \Big]\,.
\label{action_inf}
\end{equation}
As in the $R^2$-inflation model \cite{Mijic:1986iv}, the action \eqref{action_inf} produces an inflationary period compatible with the observations and the constant $\beta v^2 / (1+\alpha v^2)$ can be fixed by the amplitude of the cosmic microwave mackground power spectrum to be $10^9$ \cite{Arbey:2020ldf}. After inflation, the Unruh effect will produce the radiation energy density of the usual standard model and will also reheat the scalar field $\xi$.

The scalar field $\xi$ which appears after symmetry breaking and characterizes the variation around the minimum will evolve as in the simplest dark fluid model when neglecting the higher order terms of the potential \eqref{pot}. It can therefore replace dark energy thanks to the $V_0$ constant term, and dark matter via the mass term. The $\xi^3$ and $\xi^4$ terms have negligible effects if $v > 7\times 10^{26}$~eV. For example in galaxies, the density of dark matter gives an average value of $3\times 10^{20}$ eV for the scalar field $\xi$, so that
\begin{equation}
\begin{aligned}
\frac{m^2\xi^3}{2v}\left/\frac{m^2\xi^2}{2}\right. \simeq 5 \times 10^{-7}\,,\\
\frac{m^2\xi^4}{8v^2}\left/\frac{m^2\xi^2}{2}\right. \simeq 5\times 10^{-14}\,,
\end{aligned}
\end{equation}
and the higher order terms can be safety neglected. Therefore the action \eqref{action_sym} unifies inflation, dark energy and dark matter by assuming a unique and single scalar field.

\section{Big-Bang nucleosynthesis}
Big-Bang nucleosynthesis is generally considered to occur during radiation domination, the total energy density is mainly composed of photons, electrons, positrons, baryons, neutrinos, antineutrinos and dark matter. For a temperature of about 1 MeV, the hydrogen nuclei can fuse into helium nuclei. The reactions which produced the primordial abundances freeze out because of the Universe expansion. In the standard cosmological model, the observational measurements and the theoretical predictions obtained using the AlterBBN public code \cite{Arbey:2018zfh} are given in Table~\ref{tab::abundances}. Helium-4, helium-3 and deuterium abundances are compatible with measurements, but there is a conflict with the lithium-7 abundance. Adding a stable scalar field to the total energy density or a decaying scalar field to radiation during Big-Bang nucleosynthesis does not solve the lithium problem. However, we will can find upper limits on the energy densities of such scalar fields in order to be compatible with helium-4, helium-3 and deuterium abundance observational measurements. There is no lower limit: without scalar field the standard model is retrieved.    

\begin{table}[!h]
\begin{center}
 \begin{tabular}{|Sc|Sc|Sc|}
 \hline \rowcolor{lightgray} elements & observational measurements  & theoretical predictions \\
 \hline	$Y_p$ & $0.245 \pm 0.003$  & $0.2472 \pm 0.0006$\\
 \hline ${}^2H/H$ & $(2.569\pm 0.027) \times 10^{-5}$  & $(2.463\pm 0.074) \times 10^{-5}$\\
 \hline ${}^3He/H$ & $(1.1 \pm 0.2) \times 10^{-5}$  & $(1.03 \pm 0.03) \times 10^{-5}$\\
 \hline  ${}^7Li/H$ & $(1.6 \pm 0.3) \times 10^{-10}$ & $(5.4 \pm 0.7) \times 10^{-10}$\\
 \hline  
 \end{tabular}
 \caption{Helium-4, helium-3, deuterium and lithium-7 abundances from observational measurements \cite{Tanabashi:2018oca} and theorical predictions \cite{Arbey:2018zfh}.\label{tab::abundances}}
\end{center}
\end{table}

\subsection{Constraints on stable scalar fields}
We first consider the density of the scalar field to be a power law of the scale factor: 
\begin{equation}
\rho_{\phi} = \rho^0_{\phi} (1\text{MeV}) \times a^{-n}\,,
\end{equation}
which modifies the abundance of the elements via a modification of the Hubble rate, which is proportional to the total energy density. For example, as we can see in Figure \ref{fig::cosmo_evo}, the simplest dark fluid model evolves with $n=6$ during Big-Bang nucleosynthesis. The $\chi^2$ constraints at $95\%$ C.L. give~\cite{Arbey:2019cpf}:
\begin{equation}
\begin{aligned}
&\text{for  } n=6: \hspace{0.2cm} \rho_{\phi}\text{ (1 MeV) }  \leq 1.40 \rho_{\gamma} \text{ (1 MeV)}\,,\\
&\text{for  } n=4: \hspace{0.2cm} \rho_{\phi}\text{ (1 MeV) }  \leq 0.11 \rho_{\gamma} \text{ (1 MeV)}\,,\\
&\text{for  } n=3:\hspace{0.2cm}  \rho_{\phi}\text{ (1 MeV) }  \leq 0.005 \rho_{\gamma} \text{ (1 MeV)}\,,\\
&\text{for  } n=0: \hspace{0.2cm} \rho_{\phi}\text{ (1 MeV) }  \leq 2\times 10^{-7} \rho_{\gamma} \text{ (1 MeV) }\,. \\
\end{aligned}
\end{equation}
For $n=4$, the limit can be reinterpreted as 3 extra species of neutrinos.

\subsection{Constraints on decaying scalar fields}

\begin{figure}[h!]
\begin{minipage}[c]{0.5\textwidth}
\includegraphics[width=7.6cm]{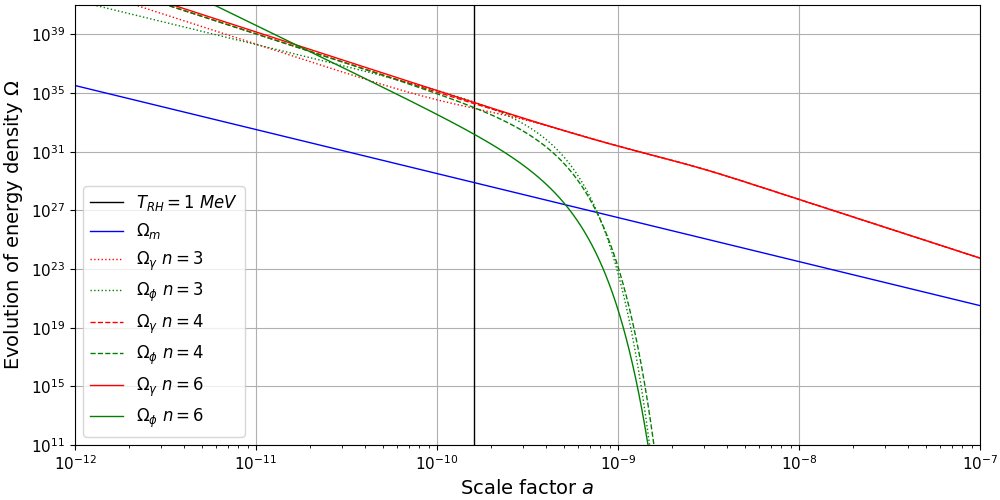}
\end{minipage}
\begin{minipage}[c]{-0.03\textwidth}
\includegraphics[width=7.6cm]{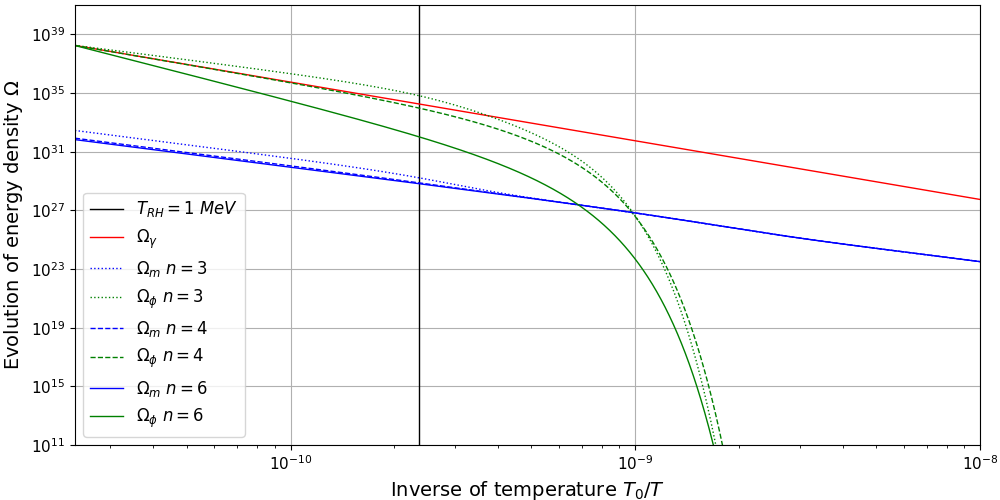}
\end{minipage}
\caption{Evolution of scalar field density (green), radiation density (red) and matter density (blue) fractions as functions of the scale factor (left) and of the inverse of temperature (right) for different values of $n$. The temperature of reheating is equal to 1 MeV and is represented by the vertical black line.\label{fig::decaying}}
\end{figure}

When the scalar field decays, it will no longer affect the Universe evolution because its density becomes negligible. If it decays after Big-Bang nucleosynthesis, the stable scalar field case is recovered. Therefore, one considers a scalar field which decays into radiation during Big-Bang nucleosynthesis. The evolution of the scalar field density is given by the Klein-Gordon equation: 
\begin{equation}
\frac{d\rho_{\phi}}{dt} = -nH\rho_{\phi}-\Gamma_{\phi}\rho_{\phi}\,,
\end{equation}
in which there is an additional decaying constant $\Gamma_{\phi} = \sqrt{4\pi^3g_{\text{eff}}\,(T_{RH})/45}T^2_{RH}/M_P$, where $T_{RH}$ is the reheating temperature. Figure \ref{fig::decaying} shows the evolution of the scalar field, which in absence of decay corresponds to $\rho_{\phi}=\rho_{\phi}^0 \times a^{-n}$. The initial value of the scalar field energy density is chosen at $T_i=10$ MeV. This value has to be adjusted to recover the baryon-to-photon ratio inferred from the cosmic microwave background. Considering the evolution of the matter, radiation and scalar field densities, it is possible to make theoretical predictions for Big-Bang nucleosynthesis. The $\chi^2$ constraints at $95\%$ C.L. give \cite{Arbey:2019cpf}:
\begin{equation}
\begin{aligned}
&\text{for   } n=6: \hspace{0.2cm} \rho_{\phi}\text{ (10 MeV) }  \leq 0.5 \rho_{\gamma} \text{ (10 MeV)}\,,\\
&\text{for   } n=4: \hspace{0.2cm} \rho_{\phi}\text{ (10 MeV) }  \leq 0.1 \rho_{\gamma} \text{ (10 MeV)}\,,\\
&\text{for   } n=3: \hspace{0.2cm} \rho_{\phi}\text{ (10 MeV) }  \leq 0.01 \left( \frac{T_{RH}}{\text{1 MeV}}\right) \rho_{\gamma} \text{ (10 MeV) }\,.\\
\end{aligned}
\end{equation}
For $n=4$, this limit is equivalent to the stable scalar field because the scalar field evolves like radiation and decays into radiation.

\section{Conclusion}

To conclude, we have presented a model that unifies not only dark matter and dark energy but also inflation. Several triple unification models have already been studied so far, but we have presented here a model in which dark matter has a behaviour similar to fuzzy dark matter i.e. an ultralight matter. Furthermore we have derived the upper limit of energy densities from Big-Bang nucleosynthesis for generic stable and decaying scalar fields. 

\bibliographystyle{h-physrev5}
\bibliography{biblio}

\end{document}